\documentclass[aps,pra,reprint,showpacs,floatfix,longbibliography]{revtex4-1}
\usepackage{graphicx, amsmath, amssymb, slashed}
\usepackage[caption=false]{subfig}
\allowdisplaybreaks[1] 
\begin{document}


\newcommand{\braket}[2]{{\left\langle #1 \middle| #2 \right\rangle}}
\newcommand{\bra}[1]{{\left\langle #1 \right|}}
\newcommand{\ket}[1]{{\left| #1 \right\rangle}}
\newcommand{\ketbra}[2]{{\left| #1 \middle\rangle \middle \langle #2 \right|}}


\title{Doubling the Success of Quantum Walk Search Using Internal-State Measurements}

\author{Kri\v{s}j\={a}nis Pr\={u}sis}
	\email{krisjanis.prusis@lu.lv}
\author{Jevg\={e}nijs Vihrovs}
	\email{jevgenijs.vihrovs@lu.lv}
\author{Thomas G.~Wong}
	\email{Currently at University of Texas at Austin, twong@cs.utexas.edu}
	\affiliation{Faculty of Computing, University of Latvia, Rai\c{n}a bulv.~19, R\=\i ga, LV-1586, Latvia}

\begin{abstract}
	In typical discrete-time quantum walk algorithms, one measures the position of the walker while ignoring its internal spin/coin state. Rather than neglecting the information in this internal state, we show that additionally measuring it doubles the success probability of many quantum spatial search algorithms. For example, this allows Grover's unstructured search problem to be solved with certainty, rather than with probability 1/2 if only the walker's position is measured, so the additional measurement yields a search algorithm that is twice as fast as without it, on average. Thus the internal state of discrete-time quantum walks holds valuable information that can be utilized to improve algorithms. Furthermore, we determine conditions for which spatial search problems on regular graphs are amenable to this doubling of the success probability, and this involves diagrammatically analyzing search using degenerate perturbation theory and deriving a useful formula for how the quantum walk acts in its reduced subspace.
\end{abstract}


\maketitle


\section{Introduction}

Discrete-time quantum walks have enjoyed much success in the development of digital quantum algorithms \cite{Ambainis2003,Kempe2003,Venegas2012}. Early on, they were shown to replicate the speedup of Grover's quantum search algorithm \cite{SKW2003}, and they optimally solve the element distinctness problem \cite{Ambainis2004}. Discrete-time quantum walks can also quickly solve any AND-OR formula \cite{Ambainis2010}, and they are universal for quantum computing \cite{Lovett2010}. To this day, new algorithmic breakthroughs are routinely developed based on discrete-time quantum walks, such as a recent scheme for speeding up backtracking algorithms \cite{Montanaro2015}.

The quantum walk can be formulated as a quantum particle hopping on a graph of $N$ vertices in superposition, where it is restricted to making local transitions defined by the edges of the graph. Thus the vertices of the graph $\{ \ket{1}, \ket{2}, \dots, \ket{N} \}$ form an orthonormal basis for $\mathbb{C}^N$, the Hilbert space of the position of the particle.

Unfortunately, these vertices are not enough to support a nontrivial quantum walk. Meyer's seminal work \cite{Meyer1996a,Meyer1996b} showed that if the graph is a homogeneous Euclidean lattice in any dimension (including the line, square lattice, cubic lattice, \textit{etc.}), the only unitary evolution for the quantum walk that exists is the trivial translation operator, times a phase.

To yield a more interesting evolution, Meyer proposed quantum walk using a $d$-level quantum particle for a quantum walk on a (regular) graph of degree $d$. Then the $d$ internal states of the particle can encode the $d$ different directions in which the particle can hop, forming an orthonormal basis for the Hilbert space $\mathbb{C}^d$. For example, for a quantum walk on the line, we use a particle with two internal degrees of freedom, which encode the particle pointing left or right. Meyer interpreted this internal degree of freedom as spin, so the particle would be a two-component spinor, and he showed that the evolution is a discretization of the Dirac equation \cite{Meyer1996a,Feynman1965}. Algorithmically, however, the internal state is often referred to as a ``coin,'' resembling flipping a coin to determine the direction of a random walk \cite{Aharonov2001,Ambainis2001,NV2000}.

Combining the position of the particle and its spin/coin internal state, the discrete-time quantum walk evolves in $\mathbb{C}^N \otimes \mathbb{C}^d$. Then the evolution can be made nontrivial. In particular, it evolves by repeated applications of the unitary operator
\begin{equation}
	\label{eq:qwalk}
	U_0 = S \cdot (I_N \otimes C_0),
\end{equation}
where $C_0$ is a ``coin flip'' that scatters the direction of the particle, and $S$ is a shift that causes the particle to hop to a neighboring vertex based on its direction. In typical quantum walk processes and algorithms, only the position of the particle is measured, while its direction is ignored (\textit{i.e.}, traced out) \cite{Venegas2012}. The position of the particle, and perhaps its spatial neighborhood \cite{Ambainis2013}, gives the solution to the computation.

In this paper, we show that the information contained in the internal state should not be neglected, that it contains valuable information that can speed up quantum algorithms. In particular, we focus on quantum search algorithms. The quantum walk formulation of Grover's algorithm is search on the complete graph, and it was shown in \cite{Wong10} to reach a success probability of $1/2$ after $\pi\sqrt{N}/2\sqrt{2}$ steps when only measuring the position of the particle. This same result holds for Shenvi, Kempe, and Whaley's search algorithm \cite{SKW2003}, which is a quantum walk on the hypercube. We show that both of these search problems can be boosted to succeed with probability $1$ by simply measuring the direction of the particle, not just its position. This doubling of the success probability holds for a variety of graphs, and we derive conditions for which spatial search problems on regular graphs are amenable to this speedup. To determine this, we analyze search using a diagrammatic approach to degenerate perturbation theory, and we derive a useful formula for how the quantum walk acts in its reduced subspace. Besides speeding up search, this suggests that the internal state contains information that can improve other discrete-time quantum walk algorithms.


\section{Search on the Complete Graph}

\subsection{Setup}

\begin{figure}
\begin{center}
	\includegraphics{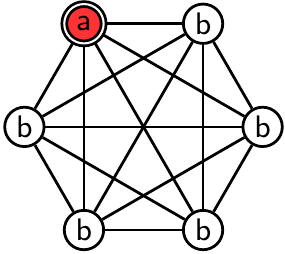}
	\caption{\label{fig:complete} The complete graph with $N = 6$ vertices. A vertex is marked, indicated by a double circle. Identically evolving vertices are identically colored and labeled.}
\end{center}
\end{figure}

We begin by specifying the specific coin and shift operators used for the quantum walk \eqref{eq:qwalk}. The usual coin is the ``Grover diffusion'' coin \cite{SKW2003}
\begin{equation}
	\label{eq:Grover_coin}
	C_0 = 2 \ketbra{s_c}{s_c} - I_d,
\end{equation}
which reflects the internal state of the particle across its equal superposition over the coin space
\[ \ket{s_c} = \frac{1}{\sqrt{d}} \sum_{i = 1}^d \ket{i}. \]
The usual shift for the quantum walk \eqref{eq:qwalk} is the flip-flop shift $S$ \cite{AKR2005}, which causes the particle to hop and then turn around (\textit{i.e.}, a particle at vertex $i$ pointing towards vertex $j$ will jump to vertex $j$ and be pointing towards vertex $i$, so $S \ket{i} \otimes \ket{i \to j} = \ket{j} \otimes \ket{j \to i}$).

Initially, we have no information as to which vertex is marked, so we guess each state with equal probability:
\begin{equation}
	\label{eq:initial}
	\ket{\psi_0} = \ket{s_v} \otimes \ket{s_c},
\end{equation}
where
\[ \ket{s_v} = \frac{1}{\sqrt{N}} \sum_{i = 1}^N \ket{i} \]
is the equal superposition over the vertices of the graph, and $\ket{s_c}$ is the equal superposition over the coin space from before. Since we have not yet introduced an oracle, walking from this initial state according to \eqref{eq:qwalk} yields no information, as expected, so $U_0 \ket{\psi_0} = \ket{\psi_0}$.

To make the quantum walk search, we introduce an oracle $R_w$ that flips the sign of the marked vertex, as in Grover's algorithm \cite{Grover1996}. That is, $R_w \ket{x} = -\ket{x}$ if $x$ is marked, and $R_w \ket{x} = \ket{x}$ if $x$ is unmarked. Querying this with each application of the quantum walk, the search algorithm is to repeatedly apply
\begin{equation}
	\label{eq:U}
	U = U_0 \cdot (R_w \otimes I_d)
\end{equation}
to the initial equal superposition state \eqref{eq:initial}. Then the number of applications of $U$ equals the number of oracle queries, which gives the runtime of the search algorithm.

\begin{figure}
\begin{center}
	\includegraphics{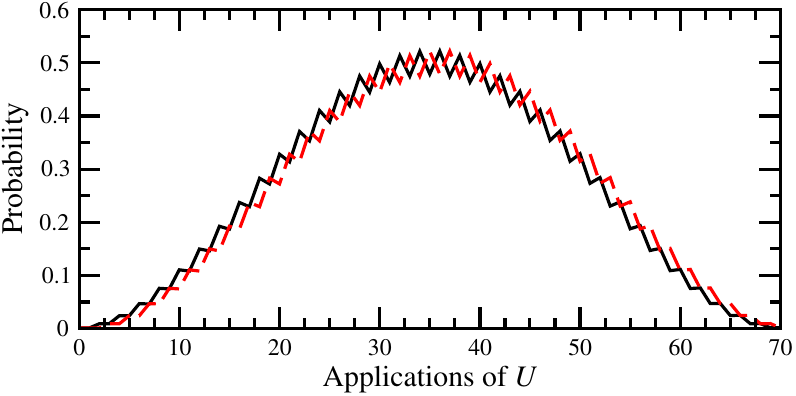}
	\caption{\label{fig:evolution_ab_ba} For search on the complete graph with $N = 1024$ vertices, the probability in $\ket{ab}$ (solid black) and $\ket{ba}$ (dashed red) as a function of the applications of the quantum walk search operator $U$.}
\end{center}
\end{figure}

Now we apply this quantum walk search algorithm to solve the unstructured search problem. This is the quantum walk analogue of Grover's algorithm \cite{Grover1996}, and it corresponds to search on the complete graph for a vertex marked by an oracle, an example of which is shown in Fig.~\ref{fig:complete}. Although this was first solved in \cite{Wong10}, we review it here and then show how measuring the internal state of the particle doubles the success probability.

Following \cite{Wong10}, the system evolves in a 3D subspace. As denoted in Fig.~\ref{fig:complete}, there are only two types of vertices, indicated by identical colors and labels: the marked $a$ vertex and the unmarked $b$ vertices. For the direction, the $a$ vertex can only point towards $b$ vertices, while the $b$ vertices can either point towards $a$ or other $b$'s. Thus the system evolves in the 3D subspace spanned by
\begin{align*}
	\ket{ab} &= \ket{a} \otimes \frac{1}{\sqrt{N-1}} \sum_b \ket{a \to b}, \\
	\ket{ba} &= \frac{1}{\sqrt{N-1}} \sum_b \ket{b} \otimes \ket{b \to a}, \\
	\ket{bb} &= \frac{1}{\sqrt{N-1}} \sum_b \ket{b} \otimes \frac{1}{\sqrt{N-2}} \sum_{b' \sim b} \ket{b \to b'}.
\end{align*}
As shown in \cite{Wong10}, after $\pi\sqrt{N}/2\sqrt{2}$ applications of $U$, the system evolves from the initial state $\ket{\psi_0}$ \eqref{eq:initial} to roughly
\begin{equation}
	\label{eq:final}
	\frac{1}{\sqrt{2}} \left( \ket{ab} - \ket{ba} \right).
\end{equation}
This state is half in $\ket{ab}$ and half in $\ket{ba}$, and that the system indeed evolves to this distribution is shown in Fig.~\ref{fig:evolution_ab_ba}, which plots $|\langle ab | U^t | \psi_0 \rangle|^2$ and $|\langle ba | U^t | \psi_0 \rangle|^2$ as functions of $t$; when $t = \pi\sqrt{1024} / 2\sqrt{2} \approx 36$, each curve reaches a probability of $1/2$.


\subsection{Doubling the Success Probability}

With the system in this final state \eqref{eq:final}, measuring only the position of the particle gives probability $1/2$ of finding it at the marked $a$ vertex because of the $\ket{ab}$ term and probability $1/2$ of finding it at a $b$ vertex because of the $\ket{ba}$ term. Typically, the analysis would end here, concluding that the algorithm succeeds with probability $1/2$. But now we make a novel observation: if the particle is at a $b$ vertex, then it is \emph{pointing towards the $a$ vertex}. Then one can measure the direction of the particle (or its internal spin/coin degree of freedom encoding the direction) to find the marked $a$ vertex. This allows us to identify the marked vertex with certainty, boosting the success probability from $1/2$ if only the position is measured to $1$ with the internal-state measurement.

More precisely, once the system is in its final state \eqref{eq:final}, the boosting procedure is: (i) Measure the position of the particle in the vertex basis $\{ \ket{1}, \ket{2}, \dots, \ket{N} \}$, so the projectors are $\ketbra{1}{1}, \ketbra{2}{2}, \dots, \ketbra{N}{N}$ (with the identity on the second tensor factor). This collapses the position of the particle to some vertex $v$. (ii) Check using the oracle if this position is marked. If yes, we are done. (iii) If not, then from \eqref{eq:final}, we know that $v$ is a $b$-type vertex $b_i$ with internal state pointing to $a$. Thus we measure the internal-state of the particle in the directional basis $\{ \ket{1}, \ket{2}, \dots, \ket{N-1} \}$, so the projectors are $\ketbra{1}{1}, \ketbra{2}{2}, \dots, \ketbra{N-1}{N-1}$. This collapses the internal state to some direction $c$, so the system is now in the state $\ket{b_i,c}$. The direction $c$ points from $b_i$ to $a$, regardless of which $b_i$ we obtained with the first measurement. (iv) Check using the oracle if the vertex in this direction from $v$ corresponds to the marked vertex, which it asymptotically will be for large $N$.

We can also consider the procedure as a single measurement in the whole $N(N-1)$-dimensional space spanned by $\ket{i,j}$, where $i = 1, \dots, N$, $j = 1, \dots, N-1$. So the projectors are $\{ \ketbra{i,j}{i,j} \}$. The first tensor factor again corresponds to the position of the particle and the second tensor factor corresponds to the directions in which a particle can point. A measurement in this computational basis collapses the state to some $\ket{v,c}$. We then check using the oracle if position $\ket{v}$ is marked, and if not, we check if the vertex in the direction of $\ket{c}$ from $\ket{v}$ is marked. So unlike typical quantum walk algorithms, we explicitly measure the internal state of the walker rather than ignore it.

Figure~\ref{fig:evolution_ab_ba} also shows that the probability in $\ket{ab}$ and $\ket{ba}$ are roughly identical throughout the entire evolution of the algorithm, not just at its runtime of $\pi\sqrt{N}/2\sqrt{2}$ applications of $U$. Thus throughout its entire evolution, the success probability can be doubled by including an internal-state measurement, since success is obtained not just from the $\ket{ab}$ term, but also the $\ket{ba}$ term. To prove this for all time, we use some analysis from \cite{Wong10}.

First we write the quantum walk search operator \eqref{eq:U} in the 3D subspace with basis $\{ \ket{ab}, \ket{ba}, \ket{bb} \}$: 
\begin{equation}
	\label{eq:U_complete}
	U =  \begin{pmatrix}
		0 & -\cos \theta & \sin \theta \\
		-1 & 0 & 0 \\
		0 & \sin \theta & \cos \theta \\
	\end{pmatrix},
\end{equation}
where
\[ \cos \theta = \frac{N-3}{N-1}, \quad \sin \theta = \frac{2\sqrt{N-2}}{N-1}. \]
Although one can explicitly work out this search operator \eqref{eq:U_complete}, we will later derive a useful formula to obtain it. To determine the evolution of the system, we decompose $U$ into its eigenvectors and eigenvalues:
\begin{gather*}
\ket{\psi_{+}} = \begin{pmatrix}
	\frac{1}{2} \sqrt{\frac{1-\cos\theta}{3+\cos\theta}} - \frac{i}{2} \\
	\frac{1}{2} \sqrt{\frac{1-\cos\theta}{3+\cos\theta}} + \frac{i}{2} \\
	\sqrt{\frac{1+\cos\theta}{3+\cos\theta}}  \\
\end{pmatrix}, \quad E_+ = e^{i\phi} \\
\ket{\psi_{-}} = \begin{pmatrix}
	\frac{1}{2} \sqrt{\frac{1-\cos\theta}{3+\cos\theta}} + \frac{i}{2} \\
	\frac{1}{2} \sqrt{\frac{1-\cos\theta}{3+\cos\theta}} - \frac{i}{2} \\
	\sqrt{\frac{1+\cos\theta}{3+\cos\theta}}  \\
\end{pmatrix}, \quad E_- = e^{-i\phi} \\
\ket{\psi_{-1}} = \begin{pmatrix}
	-\sqrt{\frac{1+\cos\theta}{3+\cos\theta}} \\
	-\sqrt{\frac{1+\cos\theta}{3+\cos\theta}} \\
	\sqrt{\frac{1-\cos\theta}{3+\cos\theta}}  \\
\end{pmatrix}, \quad E_{-1} = -1
\end{gather*}
where $\phi$ is defined such that
\[ \cos\phi = \frac{1+\cos\theta}{2}, \quad \sin\phi = \frac{\sqrt{(1-\cos\theta)(3+\cos\theta)}}{2}. \]
Note that adding the first two eigenstates yields $\ket{bb}$ for large $N$:
\[ | bb \rangle \approx \frac{1}{\sqrt{2}} \left( \left| \psi_+ \right\rangle + \left| \psi_- \right\rangle \right). \]
Since the initial state $\ket{\psi_0}$ is approximately $\ket{bb}$ for large $N$, the system after $t$ applications of $U$ is
\begin{align*}
	U^t | \psi_0 \rangle 
		&\approx \frac{1}{\sqrt{2}} \left( U^t \ket{\psi_+} + U^t \ket{\psi_-} \right) \\
		&= \frac{1}{\sqrt{2}} \left( e^{i \phi t} \ket{\psi_+} + e^{-i \phi t} \ket{\psi_-} \right).
\end{align*}
Plugging in the eigenstates and using $e^{i \phi t} + e^{-i \phi t} = 2\cos(\phi t)$ and $e^{i \phi t} - e^{-i \phi t} = 2i\sin(\phi t)$, we get
\[ U^t | \psi_0 \rangle \approx \frac{1}{\sqrt{2}} \begin{pmatrix}
	\sqrt{\frac{1-\cos\theta}{3+\cos\theta}} \cos(\phi t) + \sin(\phi t) \\
	\sqrt{\frac{1-\cos\theta}{3+\cos\theta}} \cos(\phi t) - \sin(\phi t) \\
	2 \sqrt{\frac{1+\cos\theta}{3+\cos\theta}} \cos(\phi t) \\
\end{pmatrix}. \]
Using this, we can find the difference in probability in $\ket{ab}$ and $\ket{ba}$ throughout the evolution:
\begin{align*}
	&| \langle ab | U^t | \psi_0 \rangle |^2 - | \langle ba | U^t | \psi_0 \rangle |^2 \\
	&\quad \approx 2 \sin(\phi t)\cos(\phi t) \sqrt{\frac{1-\cos\theta}{3+\cos\theta}} \\
	&\quad \approx \sin(2\phi t) \frac{1}{\sqrt{2N}} \\
	&\quad \le \frac{1}{\sqrt{2N}}.
\end{align*}
Since this difference tends towards zero as $N$ increases, the probability in $\ket{ab}$ and $\ket{ba}$ are roughly the same for large $N$ throughout the algorithm's evolution, not just at the end of the algorithm. Thus throughout the entire evolution, the success probability can be doubled by including an internal-state measurement.


\subsection{Intuition and Perturbation Theory}

\begin{figure}
\begin{center}
	\subfloat[]{
		\includegraphics{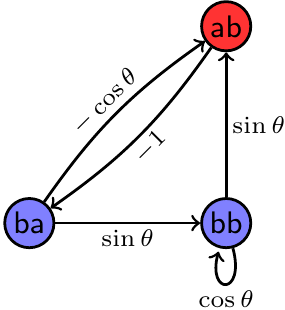}
		\label{fig:complete_diagram_U}
	} \quad
	\subfloat[] {
		\includegraphics{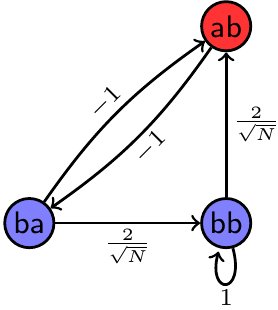}
		\label{fig:complete_diagram_U_largeN}
	}
	\caption{For the complete graph, (a) the quantum walk search operator $U$ in the $\{ \ket{ab}, \ket{ba}, \ket{bb} \}$ basis, and (b) its large $N$ approximation.}
\end{center}
\end{figure}

Although the above calculation proves that the success probability is doubled when incorporating internal-state measurements, it may not yield much intuition as to why the doubling occurs. Let us now remedy this, which will provide necessary insights to prove that the doubling of the success probability also works for search on a variety of other graphs. 

In the 3D basis, the quantum walk search operator \eqref{eq:U_complete} can be diagrammatically represented as shown in Fig.~\ref{fig:complete_diagram_U}. Although such diagrams are often used for continuous-time quantum walks \cite{Wong8}, this seems original for discrete-time quantum walks despite their seeming obviousness. For large $N$, we Taylor expand $\sin\theta \approx 2/\sqrt{N}$ and $\cos\theta \approx 1$ to yield the operator
\begin{equation}
	\label{eq:U_complete_largeN}
	U =  \begin{pmatrix}
		0 & -1 & \frac{2}{\sqrt{N}} \\
		-1 & 0 & 0 \\
		0 & \frac{2}{\sqrt{N}} & 1 \\
	\end{pmatrix},
\end{equation}
which is visualized in Fig.~\ref{fig:complete_diagram_U_largeN}. This reveals that the amplitude at $\ket{ab}$ and $\ket{ba}$ swap with minus signs with each step of the algorithm. Over the course of evolution, this causes the probability in $\ket{ab}$ and $\ket{ba}$ to be approximately equal since any influence from the other basis states ($\ket{bb}$ in this case) is averaged out over both of them. This is the intuition for $\ket{ab}$ and $\ket{ba}$ having the same probability, which furthermore allows the success probability to be doubled with an internal-state measurement.

\begin{figure}
\begin{center}
	\includegraphics{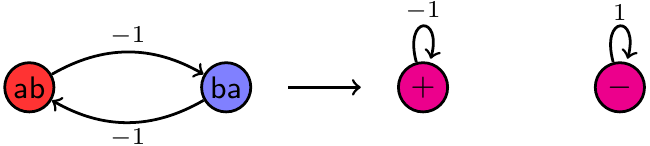}
	\caption{\label{fig:diagram_basis} By changing the basis from $\{\ket{ab},\ket{ba}\}$ to $\{\ket{+},\ket{-}\}$, when $\ket{ab}$ and $\ket{ba}$ swap their amplitudes with minus signs, this is equivalent to applying a minus sign to $\ket{+}$ and the identity to $\ket{-}$.}
\end{center}
\end{figure}

\begin{figure}
\begin{center}
	\subfloat[]{
		\includegraphics{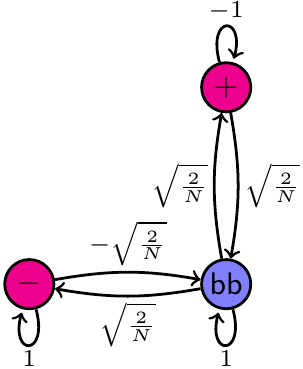}
		\label{fig:complete_diagram_U_largeN_basis}
	} \quad
	\subfloat[] {
		\includegraphics{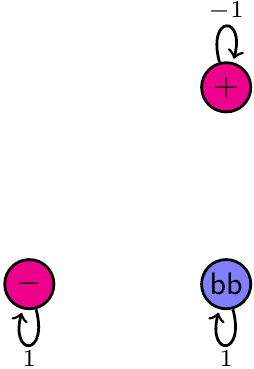}
		\label{fig:complete_diagram_U0_largeN_basis}
	}
	\caption{\label{fig:complete_diagrams_basis} For the complete graph, (a) the quantum walk search operator $U'$ for large $N$ in the $\{ \ket{+}, \ket{-}, \ket{bb} \}$ basis, and (b) its leading-order terms.}
\end{center}
\end{figure}

This intuition, that $\ket{ab}$ and $\ket{ba}$ swapping with minus signs leads to them having equal probability, can be made more rigorous by changing the basis. Instead of using $\{ \ket{ab}, \ket{ba}, \ket{bb} \}$, we use the basis
\begin{gather*}
	\ket{+} = \frac{1}{\sqrt{2}} \left( \ket{ab} + \ket{ba} \right), \\
	\ket{-} = \frac{1}{\sqrt{2}} \left( \ket{ab} - \ket{ba} \right), \\*
	\ket{bb}.
\end{gather*}
When $\ket{ab}$ and $\ket{ba}$ swap with minus signs, this change of basis causes the replacement depicted in Fig.~\ref{fig:diagram_basis}. To express the quantum walk search operator for large $N$ \eqref{eq:U_complete_largeN} in this basis, we conjugate it by
\[ T = \begin{pmatrix}
	\ket{+} & \ket{-} & \ket{bb} \\
\end{pmatrix} = \begin{pmatrix}
	\frac{1}{\sqrt{2}} & \frac{1}{\sqrt{2}} & 0 \\
	\frac{1}{\sqrt{2}} & \frac{-1}{\sqrt{2}} & 0 \\
	0 & 0 & 1 \\
\end{pmatrix}. \]
Then in the new $\{ \ket{+}, \ket{-}, \ket{bb} \}$ basis, the quantum walk search operator for large $N$ is
\[ U' = T^{-1} U T = \begin{pmatrix}
	-1 & 0 & \sqrt{\frac{2}{N}} \\
	0 & 1 & \sqrt{\frac{2}{N}} \\
	\sqrt{\frac{2}{N}} & -\sqrt{\frac{2}{N}} & 1 \\
\end{pmatrix}. \]
This can also be expressed diagrammatically, as shown in Fig.~\ref{fig:complete_diagram_U_largeN_basis}. In this diagram, note that $\ket{-}$ and $\ket{bb}$ both have self-loops with weight $1$, whereas $\ket{+}$ has a self-loop of weight $-1$. That $\ket{-}$ and $\ket{bb}$ have self-loops of equal weight is of crucial importance---as we prove next, this causes the system to evolve from $\ket{\psi_0} \approx \ket{bb}$ to $\ket{-}$. This yields the final state \eqref{eq:final}, which allows the success probability to be doubled using an internal-state measurement.

We prove this behavior using degenerate perturbation theory \cite{Griffiths2005}, which is often used to analyze continuous-time quantum walks \cite{Wong5,Wong7,Wong9,Wong11,Wong16}, but here we novelly use it in the discrete-time setting. We break the quantum walk operator into leading- and higher-order components:
\[ U' = \underbrace{\begin{pmatrix}
	-1 & 0 & 0 \\
	0 & 1 & 0 \\
	0 & 0 & 1 \\
\end{pmatrix}}_{U'_0} + \underbrace{\begin{pmatrix}
	0 & 0 & \sqrt{\frac{2}{N}} \\
	0 & 0 & \sqrt{\frac{2}{N}} \\
	\sqrt{\frac{2}{N}} & -\sqrt{\frac{2}{N}} & 0 \\
\end{pmatrix}}_{U'_1}. \]
The leading-order operator $U'_0$ is expressed diagrammatically in Fig.~\ref{fig:complete_diagram_U0_largeN_basis}. Clearly, its eigenstates are $\ket{+}$, $\ket{-}$, and $\ket{bb}$ with respective eigenvalues $-1$, $1$, and $1$. Since $\ket{-}$ and $\ket{bb}$ are degenerate, including the higher-order terms $U'_1$ causes linear combinations of them
\[ \alpha_- \ket{-} + \alpha_b \ket{bb} \]
to be approximate eigenstates of the perturbed operator $U'_0 + U'_1$. The coefficients $\alpha_-$ and $\alpha_b$ can be found by solving the eigenvalue problem
\[ \begin{pmatrix}
	U'_{--} & U'_{-b} \\
	U'_{b-} & U'_{bb} \\
\end{pmatrix} \begin{pmatrix}
	\alpha_- \\
	\alpha_b \\
\end{pmatrix} = E \begin{pmatrix}
	\alpha_- \\
	\alpha_b \\
\end{pmatrix}, \] 
where $U'_{-b} = \langle - | U'_0 + U'_1 | bb \rangle$, etc. Evaluating these matrix elements, we get
\[ \begin{pmatrix}
	1 & \sqrt{\frac{2}{N}} \\
	-\sqrt{\frac{2}{N}} & 1 \\
\end{pmatrix} \begin{pmatrix}
	\alpha_- \\
	\alpha_b \\
\end{pmatrix} = E \begin{pmatrix}
	\alpha_- \\
	\alpha_b \\
\end{pmatrix}. \] 
Solving this eigenvalue problem yields approximations for two of the eigenvectors and eigenvalues of $U'_0 + U'_1$:
\[ \ket{\psi_\pm} = \frac{1}{\sqrt{2}} \left( \mp i \ket{-} + \ket{bb} \right), \quad E_\pm = 1 \pm i\sqrt{\frac{2}{N}}. \]
This means $\ket{bb}$ is an equal superposition of these eigenstates:
\[ \ket{bb} = \frac{1}{\sqrt{2}} \left( \ket{\psi_+} + \ket{\psi_-} \right). \]
To show that the system evolves to $\ket{-}$ in time $\pi\sqrt{N}/2\sqrt{2}$, we write the eigenvalues $E_\pm$ as $e^{\pm i\sigma}$, where $\sigma \approx \sqrt{2/N}$ for large $N$. Then applying the quantum walk search operator $t$ times, the system roughly evolves to
\[ (U')^t \ket{bb} = \frac{1}{\sqrt{2}} \left( e^{i\sigma t} \ket{\psi_+} + e^{-i\sigma t} \ket{\psi_-} \right). \]
At time
\[ t = \frac{\pi}{2\sigma} \approx \frac{\pi}{2\sqrt{2}} \sqrt{N}, \]
the state becomes
\begin{align*}
	(U')^t \ket{bb} 
		&= \frac{1}{\sqrt{2}} \left( i \ket{\psi_+} - i \ket{\psi_-} \right) \\
		&= \ket{-}.
\end{align*}
This calculation using degenerate perturbation theory proves that the system evolves from $\ket{\psi_0} \approx \ket{bb}$ to $\ket{-}$, which were the states with self-loops of weight $1$ in Fig.~\ref{fig:complete_diagrams_basis}, in time $\pi\sqrt{N}/2\sqrt{2}$, in agreement with \cite{Wong10}. Relevant to this paper, this final state allows the success probability to be doubled from $1/2$ to $1$ with an internal-state measurement.


\section{Doubling Condition}

\begin{figure}
\begin{center}
	\subfloat[]{
		\includegraphics{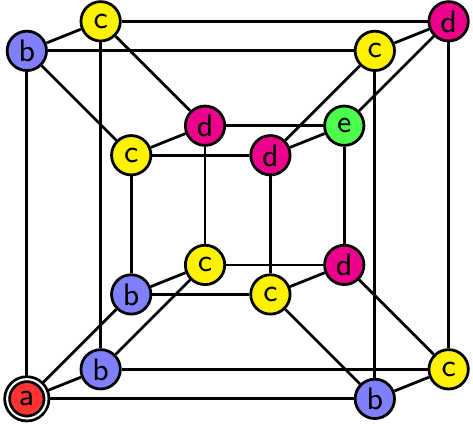}
		\label{fig:hypercube} 
	}

	\subfloat[] {
		\includegraphics{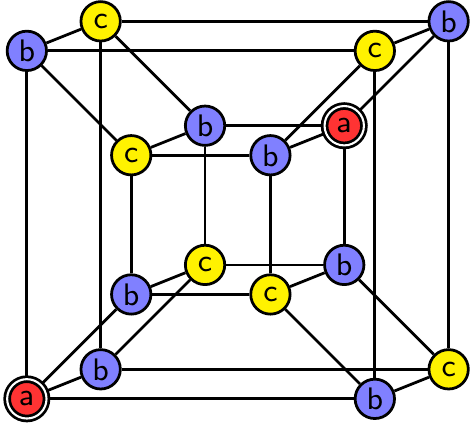}
		\label{fig:hypercube_two} 
	}
	\caption{Four-dimensional hypercube. (a) A vertex is marked, indicated by a double circle. (b) Two vertices on ``opposite ends'' are marked. Identically evolving vertices are identically colored and labeled.}
\end{center}
\end{figure}

\begin{figure}
\begin{center}
	\includegraphics{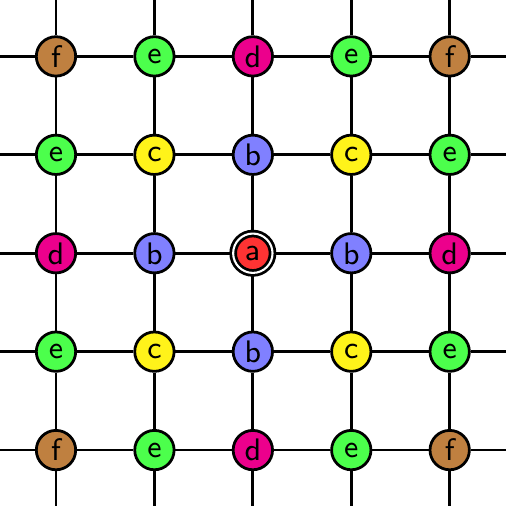}
	\caption{\label{fig:square} 2D periodic square lattice. A vertex is marked, indicated by a double circle. Identically evolving vertices are identically colored and labeled.}
\end{center}
\end{figure}

In this section, we find the condition that allows the success probability to be doubled by an internal-state measurement, assuming that marked vertices evolve identically as $a$, and they are only adjacent to one type of vertex $b$. This is satisfied by a variety of well-studied spatial search problems: With one marked-vertex, it includes the complete graph previously analyzed in Fig.~\ref{fig:complete} \cite{Wong10}, the hypercube in Fig.~\ref{fig:hypercube} \cite{SKW2003}, and arbitrary-dimensional periodic square lattices in Fig.~\ref{fig:square} \cite{AKR2005}. As an example with multiple marked vertices, it also includes the hypercube with two marked vertices on ``opposite ends'' of the hypercube, as in Fig.~\ref{fig:hypercube_two}. With this assumption, the success probability from measuring the position of the particle comes entirely from $\ket{ab}$, and we also have that $U \ket{ab} = -\ket{ba}$.

As shown in the previous section, if we additionally have that $U \ket{ba} \approx -\ket{ab}$, then $\ket{ab}$ and $\ket{ba}$ roughly swap with minus signs with each step of the search operator. This causes $\ket{ab}$ and $\ket{ba}$ to approximately evolve with equal probability, since any influence is averaged out between them. Thus to find when an internal-state measurement doubles the success probability, it suffices to find when $U \ket{ba} \approx -\ket{ab}$.

More rigorously, we first show that the initial state $\ket{\psi_0}$ \eqref{eq:initial} always has a self-loop with weight $1$, to leading-order and in the appropriate basis. For the complete graph, this corresponded to $\ket{bb}$ in Fig.~\ref{fig:complete_diagrams_basis}. Then we derive a useful formula for how $U$ \eqref{eq:U} acts on general subspace basis vectors. Using this, we then find a condition for when $U \ket{ba} \approx -\ket{ab}$, which implies that $\ket{-}$ also has a self-loop of $1$ while $\ket{+}$ has a self-loop of $-1$. Thus from degenerate perturbation theory, the system evolves from $\ket{\psi_0}$ to $\ket{-}$, and this final state enables the success probability to be doubled using an internal-state measurement.


\subsection{Initial State}

We begin by proving that the initial state $\ket{\psi_0} = \ket{s_v} \otimes \ket{s_c}$ \eqref{eq:initial} will always, to leading-order, have a self-loop of weight $1$ under the search operation $U$ \eqref{eq:U}. This is equivalent to proving that $\ket{\psi_0}$ is approximately an eigenvector of $U$ with eigenvalue $1$, for large $N$. To do this, we utilize the fact that without the oracle, $\ket{\psi_0}$ is exactly a 1-eigenvector of the quantum walk $U_0$ \eqref{eq:qwalk} alone. We prove that the addition of the oracle, which defines the search operator \eqref{eq:U}, makes a sufficiently small difference such that, to leading-order, $\ket{\psi_0}$ is also a 1-eigenvector of $U$.

For simplicity, say there is a single marked vertex $\ket{w}$. Since the oracle reflects through this state, it can be written as
\[ R_w = I_N - 2 \ketbra{w}{w}. \]
Using this, let us find the error \cite{NielsenChuang2000} between acting on $\ket{\psi_0}$ by $U$ versus $U_0$:
\begin{align*}
	\left\lVert (U - U_0) \ket{\psi_0} \right\rVert 
	&= \left\lVert (U_0 (R_w \otimes I_d) - U_0) \ket{\psi_0} \right\rVert \\
	&= \left\lVert U_0 ((R_w \otimes I_d) - I) \ket{\psi_0} \right\rVert \\
	&= \left\lVert U_0 ((I_N - 2 \ketbra{w}{w}) \otimes I_d - I) \ket{\psi_0} \right\rVert \\
	&= \left\lVert U_0 (I - 2\ketbra{w}{w} \otimes I_d - I) \ket{\psi_0} \right\rVert \\
	&= \left\lVert U_0 (-2\ketbra{w}{w} \otimes I_d) \ket{\psi_0} \right\rVert \\
	&= \left\lVert U_0 (-2\ketbra{w}{w} \otimes I_d) (\ket{s_v} \otimes \ket{s_c}) \right\rVert \\
	&= \left\lVert U_0 \left( \frac{-2}{\sqrt{N}} \ket{w} \otimes \ket{s_c} \right) \right\rVert \\
	&= \frac{2}{\sqrt{N}} \left\lVert U_0 \left( \ket{w} \otimes \ket{s_c} \right) \right\rVert \\
	&= \frac{2}{\sqrt{N}}.
\end{align*}
In the last line, note that $\ket{w} \otimes \ket{s_c}$ is a normalized quantum state. Since $U_0$ is a unitary operator, it does not change the norm of this state, so its norm is $1$. Since this error decreases with $N$, for large $N$, the initial state $\ket{\psi_0}$ is also a 1-eigenvector of $U$. Thus diagrammatically, the initial state has a self-loop of weight $1$ to leading order, although it might only manifest itself in the appropriate basis.

With $k$ marked vertices, the previous error calculation would result in $2k/\sqrt{N}$, which still tends to zero for small numbers of marked vertices, \textit{i.e.}, when $k = o(\sqrt{N})$.


\subsection{Subspace Evolution}

Since $\ket{\psi_0}$ has a self-loop of weight $1$, degenerate perturbation theory implies that the system evolves to other states with self-loops of weight $1$. Since we assume that $U \ket{ab} = -\ket{ba}$, if we additionally have that $U \ket{ba} \approx -\ket{ab}$, then $\ket{-}$ will have a self-loop of weight $1$ while $\ket{+}$ will have a self-loop of weight $-1$. Furthermore, any other states with self-loops of weight $1$ will asymptotically have no contribution to $\ket{ab}$ or $\ket{ba}$. Thus the system evolves from $\ket{\psi_0}$ to a state whose success comes from $\ket{-}$ alone. This final state allows the success probability to be doubled by an internal-state measurement. 

So finding when the success probability can be doubled is equivalent to finding when $U \ket{ba} \approx -\ket{ab}$. Since the $b$ vertices are unmarked, however, finding when $U \ket{ba} \approx -\ket{ab}$ is equivalent to finding when $U_0 \ket{ba} \approx -\ket{ab}$, since the oracle does nothing to the unmarked vertices.  To find when $U_0 \ket{ba} \approx -\ket{ab}$, we now derive a useful formula for how $U_0$ acts on subspace basis states in general, rather than only finding $U_0 \ket{ab}$ alone.

Recall that for search on the complete graph in Fig.~\ref{fig:complete}, the system evolved in a 3D subspace spanned by $\{ \ket{ab}, \ket{ba}, \ket{bb} \}$. Such a reduction is also possible for many well-studied graphs. For example, for search on the 4-dimensional hypercube in Fig.~\ref{fig:hypercube}, the system evolves in a 8-dimensional subspace spanned by $\{ \ket{ab}, \ket{ba}, \ket{bc}, \ket{cb}, \ket{cd}, \ket{dc}, \ket{de}, \ket{ed} \}$, rather than the full 64-dimensional space from the 16 vertices, each with 4 directions.

\begin{table}
\caption{\label{table:complete}Subspace basis vectors for search on the complete graph of $N$ vertices.}
\begin{ruledtabular}
\begin{tabular}{cccc}
	$\ket{xy}$ & $|x|$ & $|x \to y|$ & $|x| |x \to y|$ \\
	\colrule
	$\ket{ab}$ & $1$ & $N-1$ & $N-1$ \\
	$\ket{ba}$ & $N-1$ & $1$ & $N-1$ \\
	$\ket{bb}$ & $N-1$ & $N-2$ & $(N-1)(N-2)$ \\
\end{tabular}
\end{ruledtabular}
\end{table}

Let us generically write a basis state as $\ket{xy}$, which describes a particle in an equal superposition over the $x$ vertices pointing towards equal superpositions of $y$ vertices:
\[ \ket{xy} = \frac{1}{\sqrt{|x|}} \sum_{x} \ket{x} \otimes \frac{1}{\sqrt{|x \to y|}} \sum_{y \sim x} \ket{x \to y}. \]
Here $|x|$ denotes the number of vertices of type $x$, and $|x \to y|$ denotes the number of ways that a particle at an $x$ vertex can point towards a $y$ vertex. For example, these quantities are summarized for the complete graph in Table~\ref{table:complete}. In this table, the row for $\ket{ba}$ has $|x| = N-1$ because there are $N-1$ vertices of type $b$ (see Fig.~\ref{fig:complete}), $|x \to y| = 1$ because there is one way to point from a particular $b$ vertex to an $a$ vertex, and $|x||x\!\to\!y| = N-1$ is simply the product of these two values.

This product $|x||x \to y|$ is useful for two reasons. First, the sum of all of them (\textit{i.e.}, summing its column in Table~\ref{table:complete}) must equal $Nd$ (for a regular graph of degree $d$), so it is a convenient sanity check. Second, these products give the initial equal superposition state \eqref{eq:initial} in the subspace basis:
\begin{equation}
	\label{eq:initial_subspace}
	\ket{\psi_0} = \frac{1}{\sqrt{Nd}} \sum_{x} \sum_{y \sim x} \sqrt{|x| |x \to y|} \ket{xy}.
\end{equation}
For example, for the complete graph, the initial state is
\begin{align*}
	\ket{\psi_0} 
	&= \frac{1}{\sqrt{N(N-1)}} \Big( \sqrt{N-1} \ket{ab} + \sqrt{N-1} \ket{ba} \\
	&\hspace{1in} + \sqrt{(N-1)(N-2)} \ket{bb} \Big) \\
	&= \frac{1}{\sqrt{N}} \left( \ket{ab} + \ket{ba} + \sqrt{N-2} \ket{bb} \right).
\end{align*}

\begin{figure}
\begin{center}
	\subfloat[]{
		\includegraphics{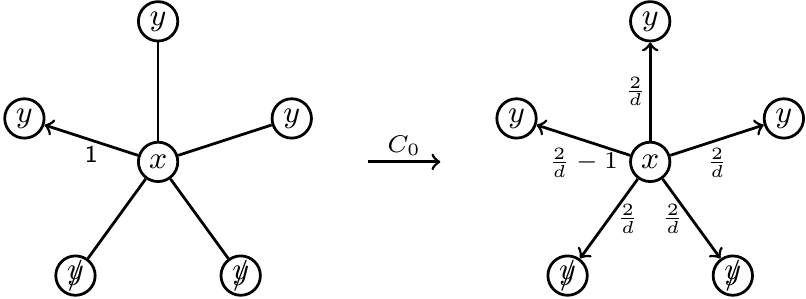}
		\label{fig:operator1}
	}

	\subfloat[]{
		\includegraphics{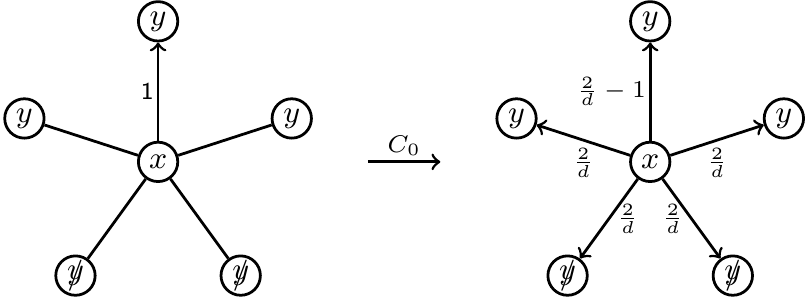}
		\label{fig:operator2}
	}

	\subfloat[]{
		\includegraphics{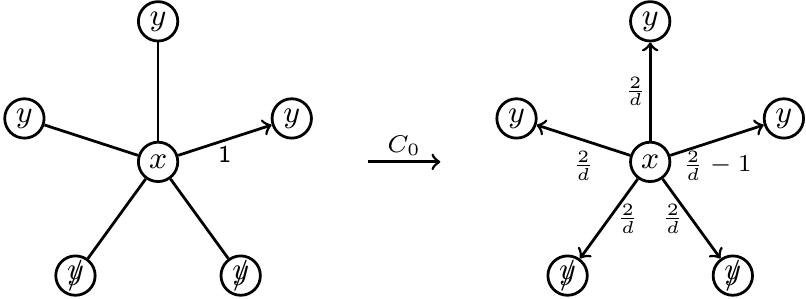}
		\label{fig:operator3}
	}

	\subfloat[]{
		\includegraphics{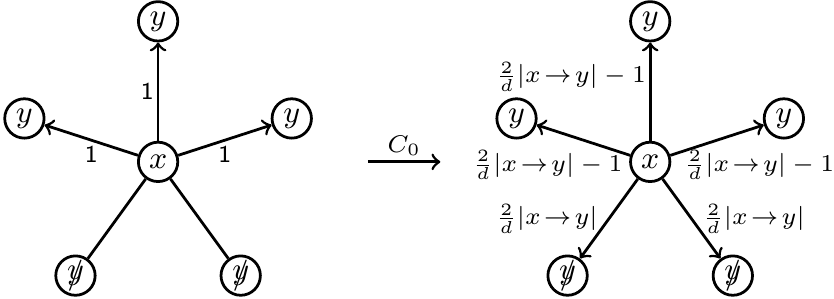}
		\label{fig:operator_sum}
	}
	\caption{For a hypothetical graph, an $x$ vertex is adjacent to three $y$ vertices and two non-$y$ vertices (denoted $\slashed{y}$). (a), (b), and (c) show the three ways that a particle at $x$ can point towards $y$, and the action of the coin $C_0$ on this internal state. (d) shows the sum of the three ways.}
\end{center}
\end{figure}

Now we want to find how the quantum walk operator $U_0$ \eqref{eq:qwalk} acts on a generic subspace basis state $\ket{xy}$. This way, we can find a condition for when $U_0 \ket{ba} \approx -\ket{ab}$, which allows the success probability to be doubled by an internal-state measurement. In particular, we will prove that $U_0$ acts according to: 
\begin{align}
	U_0 \ket{xy} 
		&= \left( \frac{2}{d} |x \to y| - 1 \right) \ket{yx} \notag \\
		&\quad + \sum_{\substack{z \sim x \\ z \ne y}} \frac{2}{d} \sqrt{|x \to y| |x \to z|} \ket{zx}. \label{eq:U0_subspace}
\end{align}
Before proving this very useful equation, let us demonstrate how to use it for the complete graph with basis vectors in Table~\ref{table:complete}. For $\ket{ba}$, for example, we have
\begin{align*}
	U_0 \ket{ba} 
		&= \left( \frac{2}{d} |b \to a| - 1 \right) \ket{ab} + \frac{2}{d} \sqrt{|b \to a| |b \to b|} \ket{bb} \\
		&= \left( \frac{2}{N-1} (1) - 1 \right) \ket{ab} + \frac{2}{N-1} \sqrt{(1)(N-2)} \ket{bb} \\
		&= -\frac{N-3}{N-1} \ket{ab} + \frac{2\sqrt{N-2}}{N-1} \ket{bb},
\end{align*}
in agreement with \eqref{eq:U_complete}. Thus from very basic properties of the vertices in Table~\ref{table:complete}, one can determine how the quantum walk acts on the subspace.

Now we prove \eqref{eq:U0_subspace}. Recall from \eqref{eq:qwalk} that $U_0 = S \cdot (I_N \otimes C_0)$. So let us first consider the action of the coin operator on $\ket{xy}$:
\[ (I_N \otimes C_0) \ket{xy} = \frac{1}{\sqrt{|x|}} \sum_{x} \ket{x} \otimes \frac{1}{\sqrt{|x \to y|}} \sum_{y \sim x} C_0 \ket{x \to y}. \]
Let us focus on the last term, $C_0 \ket{x \to y}$, where the Grover diffusion operator \eqref{eq:Grover_coin} acts on a particular $x$ vertex pointing towards a particular $y$ vertex. Hypothetically, say $x$ has structure depicted on the left side of Fig.~\ref{fig:operator1}. As the figure shows, there may be other $y$ vertices adjacent to $x$, as well as other non-$y$ vertices, which may be of different types, but for now we just write them all as $\slashed{y}$. When we apply $C_0$, we get the right side of Fig.~\ref{fig:operator1}, where the amplitude of $x$ pointing towards the original $y$ is $(2/d - 1)$, while all the other directions get an amplitude of $2/d$. Now we sum over the $y$ vertices that are adjacent to the particular $x$ vertex:
\[ \sum_{y \sim x} C_0 \ket{x \to y}. \]
In our hypothetical example, the action of $C_0$ on the other $y$ vertices is depicted in Fig.~\ref{fig:operator2} and Fig.~\ref{fig:operator3}. Then summing over $y \sim x$ is simply adding these three figures, which results in Fig.~\ref{fig:operator_sum}. In particular, each $y$ direction gets an amplitude of $(2/d - 1)$ from when $x$ points towards it, plus $2/d$ for each of the $(|x \to y| - 1)$ occurrences that $x$ points towards the other $y$ vertices. The net result of this is that each $y$ vertex gets an amplitude of
\[ \left( \frac{2}{d} - 1 \right) + \frac{2}{d} \left( |x \to y| - 1 \right) = \frac{2}{d} |x \to y| - 1. \]
For the non-$y$ vertices, each one gets an amplitude of $2/d$ for each of the $|x \to y|$ occurrences when $x$ points towards a $y$ vertex, for a net amplitude of
\[ \frac{2}{d} |x \to y|. \]
Combining these, as illustrated in Fig.~\ref{fig:operator_sum}, we have
\begin{align*}
	\sum_{y \sim x} C_0 \ket{x \to y} 
		&= \left( \frac{2}{d} |x \to y| - 1 \right) \sum_{y \sim x} \ket{x \to y} \\
		&\quad + \frac{2}{d} |x \to y| \sum_{\slashed{y} \sim x} \ket{x \to \slashed{y}}.
\end{align*}
Plugging this in, the coin acts on the basis state $\ket{xy}$ by
\begin{widetext}
\begin{align*}
	(I_N \otimes C_0) \ket{xy} 
		&= \frac{1}{\sqrt{|x|}} \sum_{x} \ket{x} \otimes \frac{1}{\sqrt{|x \to y|}} \left[ \left( \frac{2}{d} |x \to y| - 1 \right) \sum_{y \sim x} \ket{x \to y} + \frac{2}{d} |x \to y| \sum_{\slashed{y} \sim x} \ket{x \to \slashed{y}} \right] \\
		&= \left( \frac{2}{d} |x \to y| - 1 \right) \ket{xy} + \frac{1}{\sqrt{|x|}} \sum_{x} \ket{x} \otimes \frac{2}{d} \sqrt{|x \to y|} \sum_{\slashed{y} \sim x} \ket{x \to \slashed{y}}, 
\end{align*}
\end{widetext}
where we used the definition of $\ket{xy}$. For the second term, $\slashed{y}$ sums over the vertices adjacent to $x$ that are not $y$-type vertices. These can be many different types of vertices; summing over these types $z$ that are adjacent to $x$ and multiplying by $\sqrt{|x \to z|}/\sqrt{|x \to z|}$ so that we can turn them into subspace basis vectors
\[ \ket{xz} = \frac{1}{\sqrt{|x|}} \sum_{x} \ket{x} \otimes \frac{1}{\sqrt{|x \to z|}} \sum_{z \sim x} \ket{x \to z}, \]
we get
\begin{align*}
	(I_N \otimes C_0) \ket{xy} 
		&= \left( \frac{2}{d} |x \to y| - 1 \right) \ket{xy} \\
		&\quad + \sum_{\substack{z \sim x \\ z \ne y}} \frac{2}{d} \sqrt{|x \to y| |x \to z|} \ket{xz}.
\end{align*}
Finally applying the flip-flop shift $S$, $\ket{xy}$ becomes $\ket{yx}$ and $\ket{xz}$ becomes $\ket{zx}$, and we get \eqref{eq:U0_subspace}. $\hfill\square$


\subsection{Condition for Doubling Trick to Work}

Using \eqref{eq:U0_subspace}, it is straightforward to find when $U \ket{ba} = U_0 \ket{ba} \approx -\ket{ab}$. We have
\begin{align*}
	U_0 \ket{ba} 
		&= \left( \frac{2}{d} |b \to a| - 1 \right) \ket{ab} \\
		&\quad + \sum_{\substack{z \sim b \\ z \ne a}} \frac{2}{d} \sqrt{|b \to a| |b \to z|} \ket{zb}.
\end{align*}
So we want the first term to dominate as $-\ket{ab}$ and the remaining terms in the sum to be negligible. Since quantum states are normalized, however, if the first term dominates, then the remaining terms are automatically negligible. That is, it suffices to have
\[ \frac{2}{d} |b \to a| - 1 \approx -1. \]
In other words, we want $(2/d) |b \to a|$ to scale less than a constant so that $-1$ dominates. This gives the condition for $U \ket{ba} = U_0 \ket{ba} \approx -\ket{ab}$:
\begin{equation}
	\label{eq:condition}
	\frac{2}{d} |b \to a| = o(1).
\end{equation}
Since we have assumed graphs where $U \ket{ab} = -\ket{ba}$, this condition also provides that $U \ket{ba} = U_0 \ket{ba} \approx -\ket{ab}$, and so $\ket{ab}$ and $\ket{ba}$ swap back and forth with minus signs, causing them to evolve with approximately the same probability. More rigorously, we also showed that the initial state, to leading-order, has a self-loop of weight $1$, which causes the system to evolve to $\ket{-}$ rather than $\ket{+}$, allowing the success probability to be doubled by an internal-state measurement.

In the next section, we apply this condition \eqref{eq:condition} to specific spatial search problems to see if they are amenable to the trick of doubling the success probability using an internal-state measurement.


\section{Examples of Specific Graphs}

In this section, we examine specific graphs to determine if the success probability can be doubled using an internal-state measurement. All of these graphs satisfy our assumption that marked vertices evolve as $a$ and are only adjacent to identically-evolving $b$ vertices. Hence $U \ket{ab} = -\ket{ba}$. So we simply need to check the condition \eqref{eq:condition} to see if $U \ket{ba} = U_0 \ket{ba} \approx -\ket{ab}$. If so, then the doubling trick works.


\subsection{Complete Graph}

We begin with the complete graph in Fig.~\ref{fig:complete} with one marked vertex. Although we already analyzed this problem thoroughly, this serves as a quick sanity check of our condition \eqref{eq:condition}. Applying it, we get
\[ \frac{2}{d} |b \to a| = \frac{2}{N-1} (1) = o(1), \]
so the doubling trick works, as expected.


\subsection{Hypercube}

Next we consider search on the $n$-dimensional hypercube for a unique marked vertex, which was the very first quantum walk search algorithm \cite{SKW2003}. An example of this with $n = 4$ is shown in Fig.~\ref{fig:hypercube}, and it has $N = 2^n$ vertices. Using the condition \eqref{eq:condition}, we get
\[ \frac{2}{d} |b \to a| = \frac{2}{n} (1) = o(1). \]
Although $n = \log_2 N$ grows slowly with $N$, for sufficiently large $N$, the condition is satisfied. Thus the success probability can be doubled using an internal-state measurement.

More precisely, \cite{SKW2003} showed that for large $N$, the success probability reaches $1/2$ after $\pi\sqrt{N}/2\sqrt{2}$ applications of the quantum walk search operator \eqref{eq:U} when measuring the position of the particle alone. By additionally measuring the internal-state of the particle, this can be boosted to a success probability of $1$ for large $N$. This doubling is also consistent with Section 7 of \cite{HT2009}, which shows that the final state of the search algorithm is half in $\ket{ab}$ and half in $\ket{ba}$ (in our notation), and Section 2 of \cite{PGKJ2009}, which remarks that the a success probability is boosted if the coin state can be measured.

We can also generalize this to the case of two marked vertices that are on ``opposite ends'' of the hypercube, as shown in Fig.~\ref{fig:hypercube_two}. In particular, if we label the vertices as $n$-bit strings so that strings differing at a single bit are adjacent, then without loss of generalization, the marked vertices can correspond to the all $0$'s and all $1$'s strings. Then the two vertices evolve identically as $a$, and they are only adjacent to $b$ vertices. So success from the position measurement comes from $\ket{ab}$, and $U \ket{ab} = -\ket{ba}$. The condition \eqref{eq:condition} again yields
\[ \frac{2}{d} |b \to a| = \frac{2}{n} (1) = o(1), \]
and so $U \ket{ba} \approx -\ket{ab}$ for large $N$, and the doubling trick works.


\subsection{Regular Complete Bipartite Graph}

\begin{figure}
\begin{center}
	\includegraphics{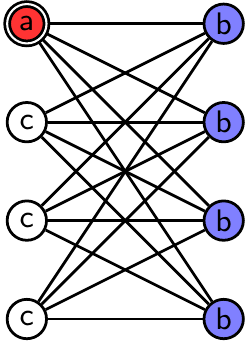}
	\caption{\label{fig:bipartite} The regular complete bipartite graph of $N = 8$ vertices. A vertex is marked, indicated by a double circle. Identically evolving vertices are identically colored and labeled.}
\end{center}
\end{figure}

For our third example, we examine search on the regular complete bipartite graph for a unique marked vertex, an example of which is shown in Fig.~\ref{fig:bipartite}. The vertices of a bipartite graph can be separated into two vertex sets, and vertices are only connected to vertices in the other set. If it is complete, then vertices are connected to \emph{all} the vertices in the other set. Taking it to be regular, the two vertex sets each have half the vertices. Then it is easy to check the condition \eqref{eq:condition}:
\[ \frac{2}{d} |b \to a| = \frac{2}{N/2} (1) = o(1), \]
So the success probability can be doubled with an internal-state measurement.


\subsection{Square, Cubic, etc. Lattices}

For the final example, we start with search on the 2D periodic square lattice for a unique marked vertex, an example of which is shown in Fig.~\ref{fig:square}. Using the condition \eqref{eq:condition}, we get
\[ \frac{2}{d} |b \to a| = \frac{2}{4} (1) = \frac{1}{2} \ne o(1). \]
The condition is not satisfied, so $U \ket{ba} = U_0 \ket{ba} \not\approx -\ket{ab}$. Thus $\ket{ab}$ and $\ket{ba}$ do not swap with minus signs, and they do not evolve with the same probability---the doubling trick does not work.

To see more explicitly why this fails, we can use \eqref{eq:U0_subspace} to show how $U$ or $U_0$ acts on $\ket{ba}$. To do this, note that a single $b$ vertex is adjacent to: the marked $a$ vertex, two $c$ vertices, and a $d$ vertex. Then using \eqref{eq:U0_subspace},
\begin{align*}
	U \ket{ba} 
		&= \left( \frac{2}{4} - 1 \right) \ket{ab} + \frac{2}{4} \sqrt{(1)(2)} \ket{cb} + \frac{2}{4} \sqrt{(1)(1)} \ket{db} \\
		&= \frac{-1}{2} \ket{ab} + \frac{1}{\sqrt{2}} \ket{cb} + \frac{1}{2} \ket{db}.
\end{align*}
Clearly, $U \ket{ba} \not\approx -\ket{ab}$, no matter how big $N$ gets.

Generalizing this to search on $D$-dimensional periodic square lattices (\textit{e.g.}, $D = 2$ is the square lattice, and $D = 3$ is the cubic lattice) for a unique marked vertex \cite{AKR2005}, the condition \eqref{eq:condition} yields
\[ \frac{2}{d} |b \to a| = \frac{2}{2D} (1) = \frac{1}{D} \stackrel{?}{=} o(1). \]
For the condition that $1/D$ to scale less than a constant to be satisfied, $D$ must scale larger than a constant. That is, $D = \omega(1)$. This implies that the doubling trick does not work for any constant-dimensional square lattice; the dimension must grow with the size of the problem.

Similar to the $D = 2$ case, we can make this more explicit using \eqref{eq:U0_subspace} to find how $U_0$ acts on $\ket{ba}$. Since $b$ is adjacent to $a$, one $d$ vertex, and $2(D-1)$ $c$ vertices, we get
\begin{align*}
	U \ket{ba} 
		&= \left( \frac{2}{2D} - 1 \right) \ket{ab} + \frac{2}{2D} \sqrt{(1)(2(D-1))} \ket{cb} \\
		&\quad + \frac{2}{2D} \sqrt{(1)(1)} \ket{db} \\
		&= \left( \frac{2}{2D} - 1 \right) \ket{ab} + \frac{\sqrt{2(D-1)}}{D} \ket{cb} + \frac{1}{D} \ket{db}.
\end{align*}
Again, for this to be approximately $-\ket{ab}$ requires that $D$ scale greater than a constant.


\section{Conclusion}

Typical discrete-time quantum walk algorithms measure the position of the randomly walking quantum particle to determine the result of a computation. So for search, one usually measures if the particle is located at the marked vertex while ignoring its internal state. We have shown, however, that the internal state contains valuable information that can contribute to the success of the search. In particular, if the particle is at a neighboring vertex and is pointing towards the marked vertex, then measuring its internal state reveals the location of the marked vertex.

For many search algorithms, this additional internal-state measurement doubles the success probability compared to a position measurement alone. Assuming that the marked vertices evolve identically as $a$ and have only one type of neighbor $b$, which encompasses many spatial search problems, we gave a condition \eqref{eq:condition} for when the doubling trick works. This involved introducing a diagrammatic analysis of discrete-time quantum walks using degenerate perturbation theory and deriving a useful formula \eqref{eq:U0_subspace} for how the quantum walk operator acts on its reduced subspace.

Thus the internal state of the particle holds valuable information that can speed up algorithms. Further work includes analyzing when the doubling trick works if the marked vertices are connected to more than one type of vertex, and how internal state measurements can improve other (non-search) discrete-time quantum walk algorithms.


\begin{acknowledgments}
	Thanks to Andris Ambainis for useful discussions. This work was supported by the European Union Seventh Framework Programme (FP7/2007-2013) under the QALGO (Grant Agreement No.~600700) project and the RAQUEL (Grant Agreement No.~323970) project, the ERC Advanced Grant MQC, and the Latvian State Research Programme NeXIT project No.~1.
\end{acknowledgments}


\bibliography{refs}

\begin{thebibliography}{28}%
\makeatletter
\providecommand \@ifxundefined [1]{%
 \@ifx{#1\undefined}
}%
\providecommand \@ifnum [1]{%
 \ifnum #1\expandafter \@firstoftwo
 \else \expandafter \@secondoftwo
 \fi
}%
\providecommand \@ifx [1]{%
 \ifx #1\expandafter \@firstoftwo
 \else \expandafter \@secondoftwo
 \fi
}%
\providecommand \natexlab [1]{#1}%
\providecommand \enquote  [1]{``#1''}%
\providecommand \bibnamefont  [1]{#1}%
\providecommand \bibfnamefont [1]{#1}%
\providecommand \citenamefont [1]{#1}%
\providecommand \href@noop [0]{\@secondoftwo}%
\providecommand \href [0]{\begingroup \@sanitize@url \@href}%
\providecommand \@href[1]{\@@startlink{#1}\@@href}%
\providecommand \@@href[1]{\endgroup#1\@@endlink}%
\providecommand \@sanitize@url [0]{\catcode `\\12\catcode `\$12\catcode
  `\&12\catcode `\#12\catcode `\^12\catcode `\_12\catcode `\%12\relax}%
\providecommand \@@startlink[1]{}%
\providecommand \@@endlink[0]{}%
\providecommand \url  [0]{\begingroup\@sanitize@url \@url }%
\providecommand \@url [1]{\endgroup\@href {#1}{\urlprefix }}%
\providecommand \urlprefix  [0]{URL }%
\providecommand \Eprint [0]{\href }%
\providecommand \doibase [0]{http://dx.doi.org/}%
\providecommand \selectlanguage [0]{\@gobble}%
\providecommand \bibinfo  [0]{\@secondoftwo}%
\providecommand \bibfield  [0]{\@secondoftwo}%
\providecommand \translation [1]{[#1]}%
\providecommand \BibitemOpen [0]{}%
\providecommand \bibitemStop [0]{}%
\providecommand \bibitemNoStop [0]{.\EOS\space}%
\providecommand \EOS [0]{\spacefactor3000\relax}%
\providecommand \BibitemShut  [1]{\csname bibitem#1\endcsname}%
\let\auto@bib@innerbib\@empty
\bibitem [{\citenamefont {Ambainis}(2003)}]{Ambainis2003}%
  \BibitemOpen
  \bibfield  {author} {\bibinfo {author} {\bibfnamefont {A.}~\bibnamefont
  {Ambainis}},\ }\bibfield  {title} {\enquote {\bibinfo {title} {Quantum walks
  and their algorithmic applications},}\ }\href {\doibase
  10.1142/S0219749903000383} {\bibfield  {journal} {\bibinfo  {journal} {Int.
  J. Quantum Inf.}\ }\textbf {\bibinfo {volume} {01}},\ \bibinfo {pages}
  {507--518} (\bibinfo {year} {2003})}\BibitemShut {NoStop}%
\bibitem [{\citenamefont {Kempe}(2003)}]{Kempe2003}%
  \BibitemOpen
  \bibfield  {author} {\bibinfo {author} {\bibfnamefont {J.}~\bibnamefont
  {Kempe}},\ }\bibfield  {title} {\enquote {\bibinfo {title} {Quantum random
  walks: An introductory overview},}\ }\href {\doibase
  10.1080/00107151031000110776} {\bibfield  {journal} {\bibinfo  {journal}
  {Contemp. Phys.}\ }\textbf {\bibinfo {volume} {44}},\ \bibinfo {pages}
  {307--327} (\bibinfo {year} {2003})}\BibitemShut {NoStop}%
\bibitem [{\citenamefont {Venegas-Andraca}(2012)}]{Venegas2012}%
  \BibitemOpen
  \bibfield  {author} {\bibinfo {author} {\bibfnamefont {S.~E.}\ \bibnamefont
  {Venegas-Andraca}},\ }\bibfield  {title} {\enquote {\bibinfo {title} {Quantum
  walks: a comprehensive review},}\ }\href {\doibase 10.1007/s11128-012-0432-5}
  {\bibfield  {journal} {\bibinfo  {journal} {Quantum Inf. Process.}\ }\textbf
  {\bibinfo {volume} {11}},\ \bibinfo {pages} {1015--1106} (\bibinfo {year}
  {2012})}\BibitemShut {NoStop}%
\bibitem [{\citenamefont {Shenvi}\ \emph {et~al.}(2003)\citenamefont {Shenvi},
  \citenamefont {Kempe},\ and\ \citenamefont {Whaley}}]{SKW2003}%
  \BibitemOpen
  \bibfield  {author} {\bibinfo {author} {\bibfnamefont {N.}~\bibnamefont
  {Shenvi}}, \bibinfo {author} {\bibfnamefont {J.}~\bibnamefont {Kempe}}, \
  and\ \bibinfo {author} {\bibfnamefont {K.~B.}\ \bibnamefont {Whaley}},\
  }\bibfield  {title} {\enquote {\bibinfo {title} {Quantum random-walk search
  algorithm},}\ }\href {\doibase 10.1103/PhysRevA.67.052307} {\bibfield
  {journal} {\bibinfo  {journal} {Phys. Rev. A}\ }\textbf {\bibinfo {volume}
  {67}},\ \bibinfo {pages} {052307} (\bibinfo {year} {2003})}\BibitemShut
  {NoStop}%
\bibitem [{\citenamefont {Ambainis}(2004)}]{Ambainis2004}%
  \BibitemOpen
  \bibfield  {author} {\bibinfo {author} {\bibfnamefont {A.}~\bibnamefont
  {Ambainis}},\ }\bibfield  {title} {\enquote {\bibinfo {title} {Quantum walk
  algorithm for element distinctness},}\ }in\ \href {\doibase
  10.1109/FOCS.2004.54} {\emph {\bibinfo {booktitle} {Proceedings of the 45th
  Annual IEEE Symposium on Foundations of Computer Science}}},\ \bibinfo
  {series and number} {FOCS '04}\ (\bibinfo  {publisher} {IEEE Computer
  Society},\ \bibinfo {year} {2004})\ pp.\ \bibinfo {pages}
  {22--31}\BibitemShut {NoStop}%
\bibitem [{\citenamefont {Ambainis}\ \emph {et~al.}(2010)\citenamefont
  {Ambainis}, \citenamefont {Childs}, \citenamefont {Reichardt}, \citenamefont
  {\v{S}palek},\ and\ \citenamefont {Zhang}}]{Ambainis2010}%
  \BibitemOpen
  \bibfield  {author} {\bibinfo {author} {\bibfnamefont {A.}~\bibnamefont
  {Ambainis}}, \bibinfo {author} {\bibfnamefont {A.~M.}\ \bibnamefont
  {Childs}}, \bibinfo {author} {\bibfnamefont {B.~W.}\ \bibnamefont
  {Reichardt}}, \bibinfo {author} {\bibfnamefont {R.}~\bibnamefont
  {\v{S}palek}}, \ and\ \bibinfo {author} {\bibfnamefont {S.}~\bibnamefont
  {Zhang}},\ }\bibfield  {title} {\enquote {\bibinfo {title} {Any {AND-OR}
  formula of size ${N}$ can be evaluated in time ${N}^{1/2+O(1)}$ on a quantum
  computer},}\ }\href {\doibase 10.1137/080712167} {\bibfield  {journal}
  {\bibinfo  {journal} {SIAM J. Comput.}\ }\textbf {\bibinfo {volume} {39}},\
  \bibinfo {pages} {2513--2530} (\bibinfo {year} {2010})}\BibitemShut {NoStop}%
\bibitem [{\citenamefont {Lovett}\ \emph {et~al.}(2010)\citenamefont {Lovett},
  \citenamefont {Cooper}, \citenamefont {Everitt}, \citenamefont {Trevers},\
  and\ \citenamefont {Kendon}}]{Lovett2010}%
  \BibitemOpen
  \bibfield  {author} {\bibinfo {author} {\bibfnamefont {N.~B.}\ \bibnamefont
  {Lovett}}, \bibinfo {author} {\bibfnamefont {S.}~\bibnamefont {Cooper}},
  \bibinfo {author} {\bibfnamefont {M.}~\bibnamefont {Everitt}}, \bibinfo
  {author} {\bibfnamefont {M.}~\bibnamefont {Trevers}}, \ and\ \bibinfo
  {author} {\bibfnamefont {V.}~\bibnamefont {Kendon}},\ }\bibfield  {title}
  {\enquote {\bibinfo {title} {Universal quantum computation using the
  discrete-time quantum walk},}\ }\href {\doibase 10.1103/PhysRevA.81.042330}
  {\bibfield  {journal} {\bibinfo  {journal} {Phys. Rev. A}\ }\textbf {\bibinfo
  {volume} {81}},\ \bibinfo {pages} {042330} (\bibinfo {year}
  {2010})}\BibitemShut {NoStop}%
\bibitem [{\citenamefont {Montanaro}(2015)}]{Montanaro2015}%
  \BibitemOpen
  \bibfield  {author} {\bibinfo {author} {\bibfnamefont {A.}~\bibnamefont
  {Montanaro}},\ }\bibfield  {title} {\enquote {\bibinfo {title} {Quantum walk
  speedup of backtracking algorithms},}\ }\href@noop {} {\bibfield  {journal}
  {\bibinfo  {journal} {{a}rXiv:1509.02374 [quant-ph]}\ } (\bibinfo {year}
  {2015})}\BibitemShut {NoStop}%
\bibitem [{\citenamefont {Meyer}(1996{\natexlab{a}})}]{Meyer1996a}%
  \BibitemOpen
  \bibfield  {author} {\bibinfo {author} {\bibfnamefont {D.~A.}\ \bibnamefont
  {Meyer}},\ }\bibfield  {title} {\enquote {\bibinfo {title} {From quantum
  cellular automata to quantum lattice gases},}\ }\href {\doibase
  10.1007/BF02199356} {\bibfield  {journal} {\bibinfo  {journal} {J. Stat.
  Phys.}\ }\textbf {\bibinfo {volume} {85}},\ \bibinfo {pages} {551--574}
  (\bibinfo {year} {1996}{\natexlab{a}})}\BibitemShut {NoStop}%
\bibitem [{\citenamefont {Meyer}(1996{\natexlab{b}})}]{Meyer1996b}%
  \BibitemOpen
  \bibfield  {author} {\bibinfo {author} {\bibfnamefont {D.~A.}\ \bibnamefont
  {Meyer}},\ }\bibfield  {title} {\enquote {\bibinfo {title} {On the absence of
  homogeneous scalar unitary cellular automata},}\ }\href {\doibase
  http://dx.doi.org/10.1016/S0375-9601(96)00745-1} {\bibfield  {journal}
  {\bibinfo  {journal} {Phys. Lett. A}\ }\textbf {\bibinfo {volume} {223}},\
  \bibinfo {pages} {337--340} (\bibinfo {year}
  {1996}{\natexlab{b}})}\BibitemShut {NoStop}%
\bibitem [{\citenamefont {Feynman}\ and\ \citenamefont
  {Hibbs}(1965)}]{Feynman1965}%
  \BibitemOpen
  \bibfield  {author} {\bibinfo {author} {\bibfnamefont {R.~P.}\ \bibnamefont
  {Feynman}}\ and\ \bibinfo {author} {\bibfnamefont {A.~R.}\ \bibnamefont
  {Hibbs}},\ }\href@noop {} {\emph {\bibinfo {title} {Quantum mechanics and
  path integrals}}},\ International series in pure and applied physics\
  (\bibinfo  {publisher} {McGraw-Hill},\ \bibinfo {address} {New York},\
  \bibinfo {year} {1965})\BibitemShut {NoStop}%
\bibitem [{\citenamefont {Aharonov}\ \emph {et~al.}(2001)\citenamefont
  {Aharonov}, \citenamefont {Ambainis}, \citenamefont {Kempe},\ and\
  \citenamefont {Vazirani}}]{Aharonov2001}%
  \BibitemOpen
  \bibfield  {author} {\bibinfo {author} {\bibfnamefont {D.}~\bibnamefont
  {Aharonov}}, \bibinfo {author} {\bibfnamefont {A.}~\bibnamefont {Ambainis}},
  \bibinfo {author} {\bibfnamefont {J.}~\bibnamefont {Kempe}}, \ and\ \bibinfo
  {author} {\bibfnamefont {U.}~\bibnamefont {Vazirani}},\ }\bibfield  {title}
  {\enquote {\bibinfo {title} {Quantum walks on graphs},}\ }in\ \href {\doibase
  10.1145/380752.380758} {\emph {\bibinfo {booktitle} {Proceedings of the 33rd
  Annual ACM Symposium on Theory of Computing}}},\ \bibinfo {series and number}
  {STOC '01}\ (\bibinfo  {publisher} {ACM},\ \bibinfo {address} {New York, NY,
  USA},\ \bibinfo {year} {2001})\ pp.\ \bibinfo {pages} {50--59}\BibitemShut
  {NoStop}%
\bibitem [{\citenamefont {Ambainis}\ \emph {et~al.}(2001)\citenamefont
  {Ambainis}, \citenamefont {Bach}, \citenamefont {Nayak}, \citenamefont
  {Vishwanath},\ and\ \citenamefont {Watrous}}]{Ambainis2001}%
  \BibitemOpen
  \bibfield  {author} {\bibinfo {author} {\bibfnamefont {A.}~\bibnamefont
  {Ambainis}}, \bibinfo {author} {\bibfnamefont {E.}~\bibnamefont {Bach}},
  \bibinfo {author} {\bibfnamefont {A.}~\bibnamefont {Nayak}}, \bibinfo
  {author} {\bibfnamefont {A.}~\bibnamefont {Vishwanath}}, \ and\ \bibinfo
  {author} {\bibfnamefont {J.}~\bibnamefont {Watrous}},\ }\bibfield  {title}
  {\enquote {\bibinfo {title} {One-dimensional quantum walks},}\ }in\ \href
  {\doibase 10.1145/380752.380757} {\emph {\bibinfo {booktitle} {Proceedings of
  the 33rd Annual ACM Symposium on Theory of Computing}}},\ \bibinfo {series
  and number} {STOC '01}\ (\bibinfo  {publisher} {ACM},\ \bibinfo {address}
  {New York, NY, USA},\ \bibinfo {year} {2001})\ pp.\ \bibinfo {pages}
  {37--49}\BibitemShut {NoStop}%
\bibitem [{\citenamefont {Nayak}\ and\ \citenamefont
  {Vishwanath}(2000)}]{NV2000}%
  \BibitemOpen
  \bibfield  {author} {\bibinfo {author} {\bibfnamefont {A.}~\bibnamefont
  {Nayak}}\ and\ \bibinfo {author} {\bibfnamefont {A.}~\bibnamefont
  {Vishwanath}},\ }\bibfield  {title} {\enquote {\bibinfo {title} {Quantum walk
  on the line},}\ }\href@noop {} {\bibfield  {journal} {\bibinfo  {journal}
  {{a}rXiv:quant-ph/0010117}\ } (\bibinfo {year} {2000})}\BibitemShut {NoStop}%
\bibitem [{\citenamefont {Ambainis}\ \emph {et~al.}(2013)\citenamefont
  {Ambainis}, \citenamefont {Ba{\v{c}}kurs}, \citenamefont {Nahimovs},
  \citenamefont {Ozols},\ and\ \citenamefont {Rivosh}}]{Ambainis2013}%
  \BibitemOpen
  \bibfield  {author} {\bibinfo {author} {\bibfnamefont {A.}~\bibnamefont
  {Ambainis}}, \bibinfo {author} {\bibfnamefont {A.}~\bibnamefont
  {Ba{\v{c}}kurs}}, \bibinfo {author} {\bibfnamefont {N.}~\bibnamefont
  {Nahimovs}}, \bibinfo {author} {\bibfnamefont {R.}~\bibnamefont {Ozols}}, \
  and\ \bibinfo {author} {\bibfnamefont {A.}~\bibnamefont {Rivosh}},\
  }\bibfield  {title} {\enquote {\bibinfo {title} {Search by quantum walks on
  two-dimensional grid without amplitude amplification},}\ }in\ \href {\doibase
  10.1007/978-3-642-35656-8_7} {\emph {\bibinfo {booktitle} {Theory of Quantum
  Computation, Communication, and Cryptography}}},\ \bibinfo {series} {Lecture
  Notes in Computer Science}, Vol.\ \bibinfo {volume} {7582},\ \bibinfo
  {editor} {edited by\ \bibinfo {editor} {\bibfnamefont {K.}~\bibnamefont
  {Iwama}}, \bibinfo {editor} {\bibfnamefont {Y.}~\bibnamefont {Kawano}}, \
  and\ \bibinfo {editor} {\bibfnamefont {M.}~\bibnamefont {Murao}}}\ (\bibinfo
  {publisher} {Springer Berlin Heidelberg},\ \bibinfo {year} {2013})\ pp.\
  \bibinfo {pages} {87--97}\BibitemShut {NoStop}%
\bibitem [{\citenamefont {Wong}(2015{\natexlab{a}})}]{Wong10}%
  \BibitemOpen
  \bibfield  {author} {\bibinfo {author} {\bibfnamefont {T.~G.}\ \bibnamefont
  {Wong}},\ }\bibfield  {title} {\enquote {\bibinfo {title} {Grover search with
  lackadaisical quantum walks},}\ }\href {\doibase
  10.1088/1751-8113/48/43/435304} {\bibfield  {journal} {\bibinfo  {journal}
  {J. Phys. A: Math. Theor.}\ }\textbf {\bibinfo {volume} {48}},\ \bibinfo
  {pages} {435304} (\bibinfo {year} {2015}{\natexlab{a}})}\BibitemShut
  {NoStop}%
\bibitem [{\citenamefont {Ambainis}\ \emph {et~al.}(2005)\citenamefont
  {Ambainis}, \citenamefont {Kempe},\ and\ \citenamefont {Rivosh}}]{AKR2005}%
  \BibitemOpen
  \bibfield  {author} {\bibinfo {author} {\bibfnamefont {A.}~\bibnamefont
  {Ambainis}}, \bibinfo {author} {\bibfnamefont {J.}~\bibnamefont {Kempe}}, \
  and\ \bibinfo {author} {\bibfnamefont {A.}~\bibnamefont {Rivosh}},\
  }\bibfield  {title} {\enquote {\bibinfo {title} {Coins make quantum walks
  faster},}\ }in\ \href@noop {} {\emph {\bibinfo {booktitle} {Proceedings of
  the 16th Annual ACM-SIAM Symposium on Discrete Algorithms}}},\ \bibinfo
  {series and number} {SODA '05}\ (\bibinfo  {publisher} {SIAM},\ \bibinfo
  {address} {Philadelphia, PA, USA},\ \bibinfo {year} {2005})\ pp.\ \bibinfo
  {pages} {1099--1108}\BibitemShut {NoStop}%
\bibitem [{\citenamefont {Grover}(1996)}]{Grover1996}%
  \BibitemOpen
  \bibfield  {author} {\bibinfo {author} {\bibfnamefont {L.~K.}\ \bibnamefont
  {Grover}},\ }\bibfield  {title} {\enquote {\bibinfo {title} {A fast quantum
  mechanical algorithm for database search},}\ }in\ \href@noop {} {\emph
  {\bibinfo {booktitle} {Proceedings of the 28th Annual ACM Symposium on Theory
  of Computing}}},\ \bibinfo {series and number} {STOC '96}\ (\bibinfo
  {publisher} {ACM},\ \bibinfo {address} {New York, NY, USA},\ \bibinfo {year}
  {1996})\ pp.\ \bibinfo {pages} {212--219}\BibitemShut {NoStop}%
\bibitem [{\citenamefont {Wong}(2015{\natexlab{b}})}]{Wong8}%
  \BibitemOpen
  \bibfield  {author} {\bibinfo {author} {\bibfnamefont {T.~G.}\ \bibnamefont
  {Wong}},\ }\bibfield  {title} {\enquote {\bibinfo {title} {Diagrammatic
  approach to quantum search},}\ }\href {\doibase 10.1007/s11128-015-0959-3}
  {\bibfield  {journal} {\bibinfo  {journal} {Quantum Inf. Process.}\ }\textbf
  {\bibinfo {volume} {14}},\ \bibinfo {pages} {1767--1775} (\bibinfo {year}
  {2015}{\natexlab{b}})}\BibitemShut {NoStop}%
\bibitem [{\citenamefont {Griffiths}(2005)}]{Griffiths2005}%
  \BibitemOpen
  \bibfield  {author} {\bibinfo {author} {\bibfnamefont {D.}~\bibnamefont
  {Griffiths}},\ }\href@noop {} {\emph {\bibinfo {title} {Introduction to
  Quantum Mechanics}}}\ (\bibinfo  {publisher} {Prentice Hall},\ \bibinfo
  {year} {2005})\BibitemShut {NoStop}%
\bibitem [{\citenamefont {Janmark}\ \emph {et~al.}(2014)\citenamefont
  {Janmark}, \citenamefont {Meyer},\ and\ \citenamefont {Wong}}]{Wong5}%
  \BibitemOpen
  \bibfield  {author} {\bibinfo {author} {\bibfnamefont {J.}~\bibnamefont
  {Janmark}}, \bibinfo {author} {\bibfnamefont {D.~A.}\ \bibnamefont {Meyer}},
  \ and\ \bibinfo {author} {\bibfnamefont {T.~G.}\ \bibnamefont {Wong}},\
  }\bibfield  {title} {\enquote {\bibinfo {title} {Global symmetry is
  unnecessary for fast quantum search},}\ }\href {\doibase
  10.1103/PhysRevLett.112.210502} {\bibfield  {journal} {\bibinfo  {journal}
  {Phys. Rev. Lett.}\ }\textbf {\bibinfo {volume} {112}},\ \bibinfo {pages}
  {210502} (\bibinfo {year} {2014})}\BibitemShut {NoStop}%
\bibitem [{\citenamefont {Meyer}\ and\ \citenamefont {Wong}(2015)}]{Wong7}%
  \BibitemOpen
  \bibfield  {author} {\bibinfo {author} {\bibfnamefont {D.~A.}\ \bibnamefont
  {Meyer}}\ and\ \bibinfo {author} {\bibfnamefont {T.~G.}\ \bibnamefont
  {Wong}},\ }\bibfield  {title} {\enquote {\bibinfo {title} {Connectivity is a
  poor indicator of fast quantum search},}\ }\href {\doibase
  10.1103/PhysRevLett.114.110503} {\bibfield  {journal} {\bibinfo  {journal}
  {Phys. Rev. Lett.}\ }\textbf {\bibinfo {volume} {114}},\ \bibinfo {pages}
  {110503} (\bibinfo {year} {2015})}\BibitemShut {NoStop}%
\bibitem [{\citenamefont {Wong}(2016)}]{Wong9}%
  \BibitemOpen
  \bibfield  {author} {\bibinfo {author} {\bibfnamefont {T.~G.}\ \bibnamefont
  {Wong}},\ }\bibfield  {title} {\enquote {\bibinfo {title} {Spatial search by
  continuous-time quantum walk with multiple marked vertices},}\ }\href
  {\doibase 10.1007/s11128-015-1239-y} {\bibfield  {journal} {\bibinfo
  {journal} {Quantum Inf. Process.}\ }\textbf {\bibinfo {volume} {15}},\
  \bibinfo {pages} {1411--1443} (\bibinfo {year} {2016})}\BibitemShut {NoStop}%
\bibitem [{\citenamefont {Wong}\ and\ \citenamefont {Ambainis}(2015)}]{Wong11}%
  \BibitemOpen
  \bibfield  {author} {\bibinfo {author} {\bibfnamefont {T.~G.}\ \bibnamefont
  {Wong}}\ and\ \bibinfo {author} {\bibfnamefont {A.}~\bibnamefont
  {Ambainis}},\ }\bibfield  {title} {\enquote {\bibinfo {title} {Quantum search
  with multiple walk steps per oracle query},}\ }\href {\doibase
  10.1103/PhysRevA.92.022338} {\bibfield  {journal} {\bibinfo  {journal} {Phys.
  Rev. A}\ }\textbf {\bibinfo {volume} {92}},\ \bibinfo {pages} {022338}
  (\bibinfo {year} {2015})}\BibitemShut {NoStop}%
\bibitem [{\citenamefont {Wong}(2015{\natexlab{c}})}]{Wong16}%
  \BibitemOpen
  \bibfield  {author} {\bibinfo {author} {\bibfnamefont {T.~G.}\ \bibnamefont
  {Wong}},\ }\bibfield  {title} {\enquote {\bibinfo {title} {Faster quantum
  walk search on a weighted graph},}\ }\href {\doibase
  10.1103/PhysRevA.92.032320} {\bibfield  {journal} {\bibinfo  {journal} {Phys.
  Rev. A}\ }\textbf {\bibinfo {volume} {92}},\ \bibinfo {pages} {032320}
  (\bibinfo {year} {2015}{\natexlab{c}})}\BibitemShut {NoStop}%
\bibitem [{\citenamefont {Nielsen}\ and\ \citenamefont
  {Chuang}(2000)}]{NielsenChuang2000}%
  \BibitemOpen
  \bibfield  {author} {\bibinfo {author} {\bibfnamefont {M.~A.}\ \bibnamefont
  {Nielsen}}\ and\ \bibinfo {author} {\bibfnamefont {I.~L.}\ \bibnamefont
  {Chuang}},\ }\href@noop {} {\emph {\bibinfo {title} {Quantum Computation and
  Quantum Information}}}\ (\bibinfo  {publisher} {Cambridge University Press},\
  \bibinfo {year} {2000})\BibitemShut {NoStop}%
\bibitem [{\citenamefont {Hein}\ and\ \citenamefont {Tanner}(2009)}]{HT2009}%
  \BibitemOpen
  \bibfield  {author} {\bibinfo {author} {\bibfnamefont {B.}~\bibnamefont
  {Hein}}\ and\ \bibinfo {author} {\bibfnamefont {G.}~\bibnamefont {Tanner}},\
  }\bibfield  {title} {\enquote {\bibinfo {title} {Quantum search algorithms on
  the hypercube},}\ }\href {\doibase 10.1088/1751-8113/42/8/085303} {\bibfield
  {journal} {\bibinfo  {journal} {Journal of Physics A: Mathematical and
  Theoretical}\ }\textbf {\bibinfo {volume} {42}},\ \bibinfo {pages} {085303}
  (\bibinfo {year} {2009})}\BibitemShut {NoStop}%
\bibitem [{\citenamefont {Poto\ifmmode~\check{c}\else \v{c}\fi{}ek}\ \emph
  {et~al.}(2009)\citenamefont {Poto\ifmmode~\check{c}\else \v{c}\fi{}ek},
  \citenamefont {G\'abris}, \citenamefont {Kiss},\ and\ \citenamefont
  {Jex}}]{PGKJ2009}%
  \BibitemOpen
  \bibfield  {author} {\bibinfo {author} {\bibfnamefont {V.}~\bibnamefont
  {Poto\ifmmode~\check{c}\else \v{c}\fi{}ek}}, \bibinfo {author} {\bibfnamefont
  {A.}~\bibnamefont {G\'abris}}, \bibinfo {author} {\bibfnamefont
  {T.}~\bibnamefont {Kiss}}, \ and\ \bibinfo {author} {\bibfnamefont
  {I.}~\bibnamefont {Jex}},\ }\bibfield  {title} {\enquote {\bibinfo {title}
  {Optimized quantum random-walk search algorithms on the hypercube},}\ }\href
  {\doibase 10.1103/PhysRevA.79.012325} {\bibfield  {journal} {\bibinfo
  {journal} {Phys. Rev. A}\ }\textbf {\bibinfo {volume} {79}},\ \bibinfo
  {pages} {012325} (\bibinfo {year} {2009})}\BibitemShut {NoStop}%
\end{thebibliography}%

\end{document}